# Observation of Quantized Hall Effect and Shubnikov-de Hass Oscillations in Highly Doped $Bi_2Se_3$: Evidence for Layered Transport of Bulk Carriers

Helin Cao[1, 2], Jifa Tian[1, 2], Ireneusz Miotkowski[1], Tian Shen[1, 3], Jiuning Hu[2, 4], Shan Qiao[5], Yong P. Chen[1, 2, 4, *]

[1]Department of Physics, Purdue University, West Lafayette, IN 47907 USA

[2]Birck Nanotechnology Center, Purdue University, West Lafayette, IN 47907 USA

[3]Physical Measurement Laboratory, National Institute of Standard and Technology, Gaithersburg, MD 20899 USA

[4]School of Electrical and Computer Engineering, Purdue University, West Lafayette, IN 47907 USA

[5]Department of Physics, Fudan University, Shanghai 200433, People's Republic of China

[*]To whom correspondence should be addressed: yongchen@purdue.edu

**$Bi_2Se_3$ is an important semiconductor thermoelectric material and a prototype topological insulator. Here we report observation of Shubnikov-de Hass (SdH) oscillations accompanied by *quantized Hall resistances* ($R_{xy}$) in highly-doped *n*-type $Bi_2Se_3$ with bulk carrier concentrations of few $10^{19}$ cm$^{-3}$. Measurements under tilted magnetic fields show that the magnetotransport is 2D-like, where only the c-axis component of the magnetic field controls the Landau level formation. The quantized step size in $1/R_{xy}$ is found to scale with the sample thickness, and average $\sim e^2/h$ per quintuple layer (QL). We show that the observed magnetotransport features do not come from the sample surface, but arise from the bulk of the sample acting as many parallel 2D electron systems to give a multilayered quantum Hall effect. Besides revealing a new electronic property of $Bi_2Se_3$, our finding also has important implications for electronic transport studies of topological insulator materials.**

Bi$_2$Se$_3$ has been extensively studied for decades and is well known as an excellent thermoelectric material [1]. Structurally, Bi$_2$Se$_3$ is made of van der Waals coupled, stacking "quintuple layers" (QL). Electronically, Bi$_2$Se$_3$ is a narrow-gap semiconductor with a band gap of ~0.3 eV. Recently, Bi$_2$Se$_3$ has also attracted strong interests as a "proto-type" of a newly identified class of electronic materials known as three dimensional (3D) topological insulators (TI) [2,3], which feature nontrivial, topologically protected metallic (gap-less) surface states [4-8]. Resulting from the interplay between the topology of the electronic band structure and strong spin-orbit coupling in the bulk, the surface state of TI gives rise to 2D Dirac fermions with spin-momentum locking and suppressed back scattering, promising a host of novel physics and devices [4-8]. The TI surface state of Bi$_2$Se$_3$ has been directly revealed in surface sensitive measurements including angle-resolved photoemission spectroscopy (ARPES) [2] and scanning tunneling microscopy (STM) [9]. Transport signatures of the TI surface state in Bi$_2$Se$_3$ have also been studied, for example, in the measurements of Aharonov-Bohm interference (in nanoribbons) [10], Shubnikov-de Hass (SdH) oscillations [11], electric field effect [12], weak-anti localization [13], etc. Typical bulk crystals of Bi$_2$Se$_3$ contain unintentional n-type doping due to Se vacancies [14-18]. Samples of low or compensated bulk doping [11] are generally preferred for studying TI surface state transport, which would otherwise be overwhelmed by the bulk conduction. In this paper, we report magnetotransport measurements performed in *highly* doped *n*-type Bi$_2$Se$_3$, where the 3D bulk carriers are expected to dominate the transport. We discover that our samples display 2D-like SdH oscillations, accompanied by a quantized Hall effect (QHE). The quantized Hall resistances ($R_{xy}$) take values $\sim \frac{1}{NZ}\frac{h}{e^2}$, where N is the Landau level index and Z is the sample thickness in the unit of number of stacked QLs making up the bulk. The observed 2D-like magnetotransport and QHE are attributed to the *bulk* of the sample behaving as many parallel 2D electron systems and *not* due to the sample surface. Such phenomena have not been found in previous experiments on Bi$_2$Se$_3$ with lower bulk doping than ours. Our results suggest a new bulk electronic state, the precise nature of which is yet to be determined, may form in Bi$_2$Se$_3$ in highly electron-doped regime. Our findings are also relevant for the transport studies of TI (focusing on the surface state), where a key task (and challenge) is to distinguish and separate surface state transport from the bulk transport, and a good understanding of bulk electronic properties is generally valuable.

High quality Bi$_2$Se$_3$ single crystals have been synthesized by the Bridgman technique [19]. Exfoliated thin flakes and thicker cleaved bulk crystals have been fabricated into devices with quasi-Hall-bar geometry. The cleaved/exfoliated sample surface, on which the electrodes are placed, is perpendicular to the c-axis (along [111] direction of Bi$_2$Se$_3$). Transport measurements (down to temperatures ~ 450 mK) have been performed on several devices, yielding similar results. Data from one sample ("A") are presented in Figs. 1 – 3.

Fig. 1a shows the temperature dependence of the resistivity of the sample, displaying a metallic behavior. In the following we focus on the magnetotransport data. Fig. 1b shows four-terminal longitudinal resistance R$_{xx}$ and Hall resistance R$_{xy}$ as functions of perpendicular magnetic field (B, applied along the c-axis) at 450 mK. The sample is n-type, with a bulk carrier density $n_{3D}^{Hall} = 4.7 \times 10^{19}$ cm$^{-3}$ as determined from the low B Hall slope. At higher B, R$_{xx}$ oscillates periodically in 1/B (Fourier analysis of the oscillations reveals a single frequency B$_F$ = 162 T). This can be interpreted as SdH oscillations due to the formation of Landau levels (LL) in high magnetic field. Furthermore, we observe developing quantized plateaus in R$_{xy}$, accompanying the minima in R$_{xx}$. Such plateaus will be interpreted as a quantized Hall effect (QHE) due to parallel 2D electron systems in the bulk of the sample, as discussed later in this paper. We find that B$_F$/B, where B is the magnetic field position of each R$_{xx}$ minimum, is very close to an integer, which we assign as the corresponding LL index N. The upper inset of Fig. 1b shows B$_F$/B vs. N. The data can be fitted to a straight line with slope 1 and intercept ~0 in the horizontal (N) axis.

The magnetotransport is found to be 2D-like by performing measurements under tilted magnetic fields (Fig. 2). The tilt angle θ between B and c-axis (schematically shown in Fig. 2c inset) can be varied from 0 to 90°, and we find the SdH oscillations and QHE are no longer observable for θ > 60° within the experimental resolution. We have found that the B position of R$_{xx}$ minimum (or R$_{xy}$ plateau) corresponding to each LL has a 1/cos(θ) dependence on the tilt angle θ. A representative example is shown for the B position of the 13$^{th}$ LL minima (B$_{13}$) as a function of θ in Fig. 2a inset, where the data can be well fitted by B$_{13}$(θ = 0)/cos(θ) (black solid line). Magnetotransport dependent on Landau levels of 2D carriers should be controlled only by the perpendicular component, $B_\perp = B \cdot \cos\theta$, of the magnetic field. We plot R$_{xx}$ (R$_{xy}$) against B$_\perp$ in Fig. 2b (c), showing that for each LL, the corresponding minimum in $R_{xx}$ (the plateau in $R_{xy}$) occurs at the same $B_\perp$ for different tilt angles. Our data demonstrate that the carriers giving rise to the observed SdH oscillations and QHE exhibit 2D-like transport, even though such carriers reside in the bulk of a 3D sample, as shown below.

Fig. 3a displays the temperature (T) dependence of SdH oscillations in $\Delta R_{xx}$ (calculated from $R_{xx}$ by subtracting a polynomial fit to the background). The amplitude of the SdH oscillations decreases with increasing T, and the oscillations are not observed for T > 50 K. We fit the T-dependence of SdH oscillation amplitude $\Delta R_{xx}$ to Lifshitz-Kosevich theory [20]

$$\Delta R_{xx}(T, B) \propto \frac{\alpha T/\Delta E_N(B)}{\sinh(\alpha T/\Delta E_N(B))} \cdot e^{-\alpha T_D/\Delta E_N(B)} \qquad (1)$$

In Eq. 1, $\Delta E_N$ and $T_D$ are the fitting parameters, and B is the magnetic field position of the $N^{th}$ minimum in $R_{xx}$. $\Delta E_N(B) = heB/2\pi m^*$ is the energy gap between $N^{th}$ and $(N+1)^{th}$ LL, where m* is the effective mass of the carriers, $e$ is electron charge, and $h$ is Planck constant. $T_D = \frac{h}{4\pi^2 \tau k_B}$ is the Dingle temperature, where $\tau$ is the quantum lifetime of carriers due to scattering, $k_B$ is Boltzmann's constant, and $\alpha = 2\pi^2 k_B$. We plot the relative amplitude $\Delta R/R_B$ as a function of T for the $10^{th}$ LL in Fig. 3b, where $R_B = \Delta R$ (T = 450 mK). The solid line shows the fit to function $\frac{\alpha T/\Delta E_N(B)}{\sinh(\alpha T/\Delta E_N(B))}$. $\Delta E_N$ calculated from such fitting for different LLs are plotted in Fig. 3b inset, and found to have an approximately linear dependence on B. The slope of a linear fitting (dashed line) yields m*~ 0.14$m_e$ ($m_e = 9.1 \times 10^{-31}$ kg is the electron mass), consistent with previous measurements [14-18]. The Dingle temperature $T_D$ is found to be 25 K from the slope in the semilog plot of D=$\Delta$R·B·sinh($\alpha$T/$\Delta E_N$) vs. 1/B (Fig. 3c) at T=450 mK. From $T_D$ we extract the carrier lifetime $\tau = \frac{h}{4\pi^2 T_D k_B} \sim 5 \times 10^{-14}$ s, which is shorter (by a factor of 2-3) than the $\tau$ measured previously in samples of lower doping (carrier density ~5×10$^{18}$/cm$^3$) [18] than ours. This is qualitatively consistent with the larger amount of impurities (Se vacancies) giving rise to more frequent carrier scattering in our samples.

In Fig. 4a, we plot 1/$R_{xy}$ divided by the number (Z) of QLs in sample "A" as a function of LL index N, where Z is calculated from the measured thickness (given the thickness of one QL is 1 nm [21]). It shows the step size, $\Delta(1/R_{xy})$, between the plateaus in 1/$R_{xy}$ is approximately constant for different LLs and ~1.2$e^2/h$ per QL. We have measured several samples with different thicknesses (ranging from 60 nm to 0.853 mm), and calculated Z' = $\Delta(1/R_{xy})/(e^2/h)$, where $\Delta(1/R_{xy})$ is the step size between adjacent quantized plateaus in $1/R_{xy}$. We plot Z' against Z in Fig. 4b, and find Z'~Z. It shows the step size in $1/R_{xy}$ scales with the sample thickness (over 4 orders of magnitude), and averages ~$e^2/h$ per QL. In fact, we have found that the $R_{xy}$ plateau values for all the samples measured (with $n$ range from ~$3 \times 10^{19}$ cm$^{-3}$ to ~$6 \times 10^{19}$ cm$^{-3}$) to be $R_{xy} = \frac{1}{\eta \cdot N \cdot Z} \frac{h}{e^2}$, where $\eta = 1 \pm 0.2$ and N is the corresponding LL index.

We interpret our observed QHE (and 2D SdH) as due to transport through many parallel 2D conduction channels (each acting as a 2D electron system, 2DES) making up the bulk of the highly doped n-type $Bi_2Se_3$. Similar QHE (sometimes called a "bulk quantum Hall effect"), where quantized $R_{xy}$ values inversely scale with the sample thickness, has been previously observed in a number of anisotropic, layered 3D electronic materials, e.g., GaAs/AlGaAs multi-quantum-wells [22], Bechgaard salts [23-25] and $Mo_4O_{11}$ [26]. Such bulk QHE is generally attributed to parallel 2D conduction channels, each made from one or few stacking layers (in our case, the QLs, as discussed below).

A 2D carrier density for each parallel 2D conduction channel can be extracted from the SdH oscillations as $n_{2D}^{SdH} = \frac{g e B_F}{h}$, where we take LL degeneracy g=2 for spin-unresolved LLs as consistent with the LL energy gaps extracted in our sample (Fig. 3b). Comparing $n_{2D}^{SdH}$ and the bulk carrier density ($n_{3D}^{Hall}$) measured from the Hall effect allows us to obtain an effective thickness ($t_{2D}$) per 2D conduction channel as $t_{2D} = n_{2D}^{SdH}/n_{3D}^{Hall}$. For sample A presented above, we find $n_{2D}^{SdH} = 7.8 \times 10^{12}$ cm$^{-2}$ and $t_{2D}$ =1.7 nm. Other samples measured give comparable values of $t_{2D}$ ~2 nm on average, close to the thickness of 2 QLs (the small variation of $t_{2D}$ among samples could be related to factors such as uncertainties in the thickness measurement, sample defects or electronic inhomogeneity). This means there are ~Z/2 parallel 2DES conduction channels in our sample, consistent with our observed $\Delta(1/R_{xy}) \sim Z \cdot \frac{e^2}{h} = \frac{Z}{2} \cdot \frac{2e^2}{h}$ (i.e., ~$2e^2/h$ for every 2 QLs), as QHE in a 2DES with LL degeneracy g=2 typically has a plateau step size in $1/R_{xy}$ of $2e^2/h$. More work is needed to understand what determines the thickness of the effective 2D conduction channels contributing in parallel to the observed QHE in our samples.

While ARPES measurements have revealed the existence of both the topological surface states [16,27] and the non-topological surface 2DES [27] in high doping $Bi_2Se_3$ samples, we do not believe that our observed "2D-like" SdH oscillations and QHE can be attributed to any types of surface conduction channels. The observation that $\Delta(1/R_{xy})$ scales with the sample *thickness* indicates the effect has a *bulk* origin, rather than a surface origin. Because of the high bulk doping in our $Bi_2Se_3$ samples, the conduction is dominated by the bulk, and the contribution of the surface to the transport is expected to be negligible. This is in contrast to the recent experiment [11], where 2D magnetotransport (including QHE features) associated with the surface state has been reported in $Bi_2Se_3$ with very low bulk carrier densities ($n < 10^{17}$ cm$^{-3}$, about 3 orders of magnitude lower than ours). We have further ruled out TI surface states as responsible for the observed QHE and SdH oscillations in our samples by the following two control experiments. 1) We have deposited a thin layer (1 nm) of magnetic impurities (Ni) on the top surface of one sample. Such magnetic impurities have been suggested to strongly disrupt the TI surface states by breaking the time-reversal symmetry [28, 29]. However, we have not observed any noticeable effect of such deposited Ni on the magnetotransport features (SdH and QHE) discussed above. 2) We have also measured $R_{xx}$ of a sample with electrodes placed on its side surface (with B still perpendicular to the surface, as well as to both c-axis and the current). While the TI surface state is supposed to exist on all the surfaces, we no longer observe any SdH oscillations (up to B = 18 T) in such side surface transport. In addition, our observed SdH oscillations give zero intercept in the LL fan diagram (Fig. 1b inset), in contrast to the finite intercept associated with Dirac fermions arising from TI surface states [30].

Several earlier experiments [14-18] have systematically studied the magnetotransport in $Bi_2Se_3$ samples with *n* ranging from ~$10^{17}$ cm$^{-3}$ to ~$10^{19}$ cm$^{-3}$. The SdH oscillations were observed for all B directions, and consistent with bulk carriers with a 3D Fermi surface (FS) that becomes moderately elongated along c-axis as *n* increases (with anisotropy factor < 2) [14-18]. No QHE was reported in these experiments. Our results, along with those from earlier experiments [11, 14-18] on less-doped $Bi_2Se_3$, demonstrate the rich electronic properties of $Bi_2Se_3$ and a remarkable "*dimensional crossover*" in its magnetotransport behavior (from 2D to 3D then back to 2D) in different doping regimes of bulk carrier densities (*n*): at very low *n* (<~$10^{17}$ cm$^{-3}$) with diminishing bulk conduction, a TI is manifested with 2D surface state transport [11]; at intermediate *n* (~$10^{17}$ – ~$10^{19}$ cm$^{-3}$), the transport is dominated by the bulk, displaying 3D SdH [14-18]; at very high *n* (> ~$3\times10^{19}$ cm$^{-3}$) as in our work, the bulk-dominated transport displays 2D-like SdH and QHE consistent with the bulk acting as many parallel 2D conduction channels.

The exact physical mechanisms that may drive highly-doped $Bi_2Se_3$ into displaying layered bulk QHE are not yet clear. We note that the calculated band structure for $Bi_2Se_3$ [3] shows that the bulk FS should remain 3D at the level of *n* in our samples, although this not yet confirmed experimentally and further work is needed to understand bulk electronic state of $Bi_2Se_3$ at very high doping (including possible effects of impurities as well as magnetic fields). We also note that recent ARPES measurements [27,31] have revealed increased deformation and trigonal warping of the *bulk* conduction band (BCB) FS at such high *n*. More work is needed to understand whether the increased warping (and nesting) of BCB FS may drive instabilities (in the bulk) toward spin or charge density waves, as such density waves have been shown to be relevant for bulk QHE in several previous examples [23-26].

The values of $R_{xx}$ minima (eg. Fig. 1 and 2) accompanying our quantized Hall ($R_{xy}$) plateaux are not close to zero as in usual QHE, indicating there is still dissipation in our samples at the "bulk" QH states. One possible source of such dissipation may be incomplete bulk localization related to the doping impurities or the coupling between the parallel 2D conducting layers in the bulk [32,33]. Another possible source of dissipation may be the non-chiral TI surface state on the side surfaces (parallel to the magnetic field), coexisting with the chiral edge states (which stack up to form a chiral surface state with many interesting predicted properties [32, 33]) in the multilayered "bulk" QHE system [34]. Further studies are needed to investigate the relative importance of different possible sources of dissipation in our experiments. We point out that the co-existence of the chiral surface state with the non-chiral TI surface state is a novel situation not present in previously studied bulk QHE systems, and the $R_{xx}$ in the bulk QHE may potentially provide insights about the side TI surface state itself.

Our work also has important implications for transport studies of 3D TIs by providing a general caution that observing 2D transport behavior alone does not necessarily indicate TI surface state. Given the challenges associated with extracting 2D surface state transport from measuring a 3D sample whose bulk is often conducting, it is important to know other (non TI-surface-state) sources of carriers in the TI materials that could also give 2D transport features. For example, it has been shown that band-bending can generate a trivial (non-topological) 2DES on $Bi_2Se_3$ surfaces, coexisting with the 2D

Dirac fermions from the topological surface state [27]. The measurements presented here, performed on highly-doped n-type $Bi_2Se_3$, reveal that even the bulk carriers can display layered, 2D-like transport.

We acknowledge partial support from a Purdue Birck/MIND Center seed grant and Intel. The magnetotransport measurements were performed at National High Magnetic Field Laboratory (NHMFL), which is jointly supported by the National Science Foundation (DMR0654118) and the State of Florida. We thank E. Palm, L. Engel and Z. Jiang for experimental help. We also acknowledge discussions with J. P. Hu, Z. Fang and L. Balicas.

**Figures**

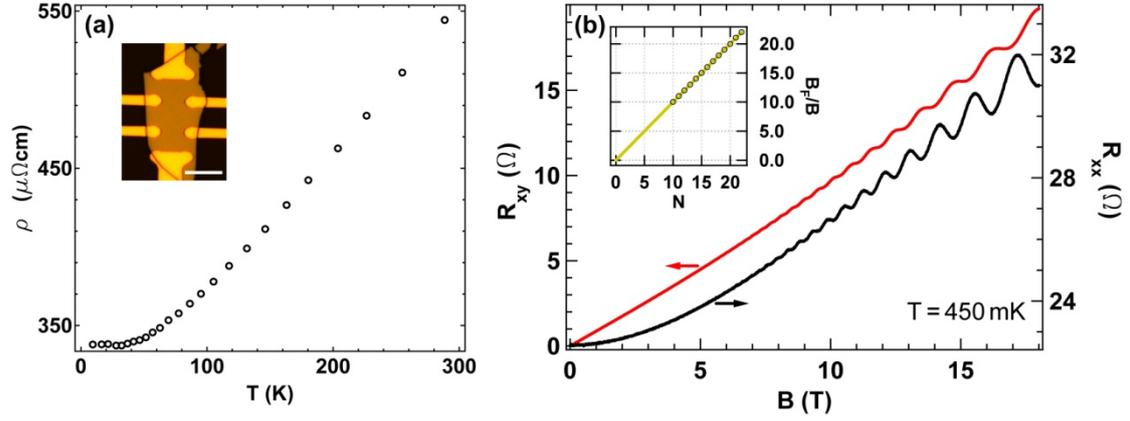

FIG. 1 (a) Temperature dependence of resistivity ρ in sample "A", displaying a metallic behavior. Inset shows the optical image of the sample (scale bar is 10 μm). (b) Hall resistance ($R_{xy}$) and four terminal longitudinal resistance ($R_{xx}$) of sample "A" as functions of perpendicular magnetic field (B) applied along c-axis (perpendicular to sample surface). Data presented were measured at T = 450 mK. Inset shows $B_F/B$ at the magnetic field (B) positions of the observed minima in $R_{xx}(B)$, plotted against assigned LL index N. The solid line is a linear fit with slope 1±0.002 and N-axis intercept 0±0.03.

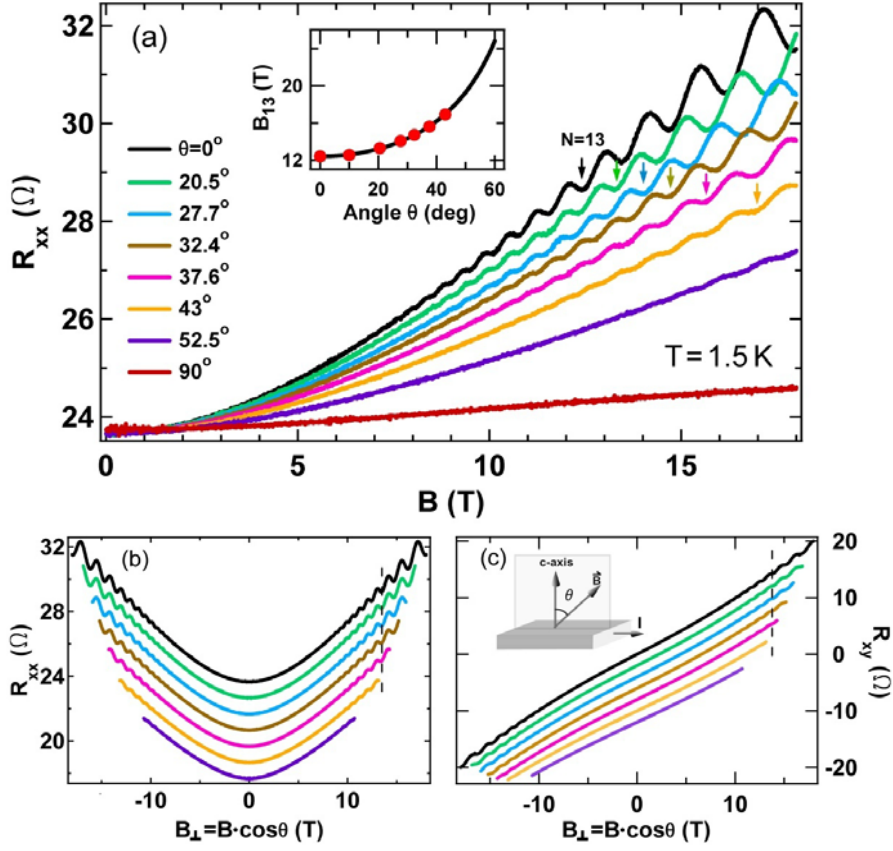

FIG. 2. (a) $R_{xx}$ as a function of B measured at various tilt angles (θ). The minima (at $B = B_{13}$) in $R_{xx}$ corresponding to N = 13 are labeled with arrows. The inset shows $B_{13}$ varies with θ as $B_{13}(\theta = 0)/\cos(\theta)$ (black solid line), indicating a 2D transport behavior from the charge carriers. (b, c) $R_{xx}$ and $R_{xy}$ plotted as functions of $B_\perp = B\cdot\cos(\theta)$, respectively. The inset in Fig. 2c shows the schematic of the sample in a tilted magnetic field (B).

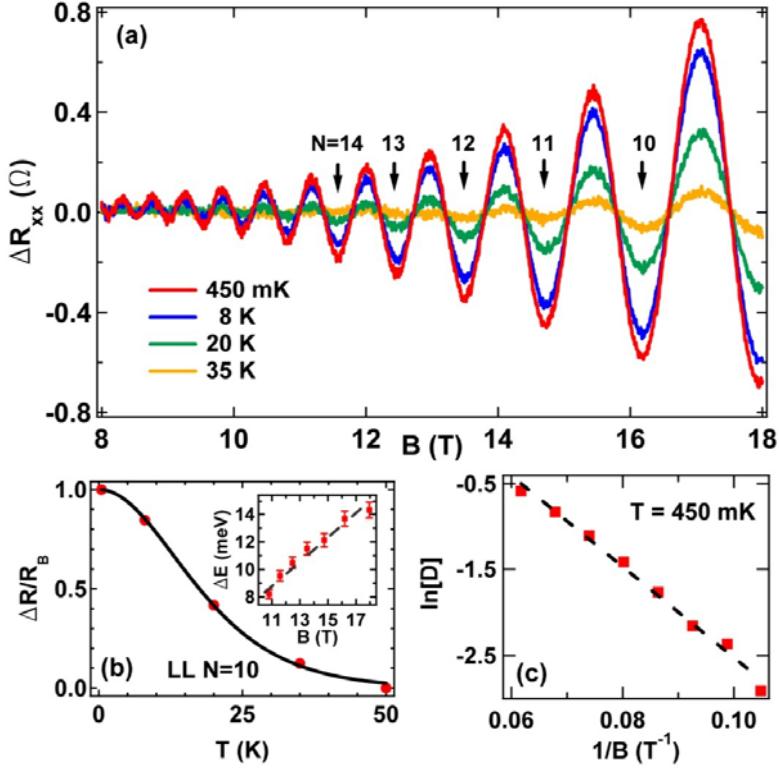

FIG. 3. (a) $\Delta R_{xx}(B)$, extracted from $R_{xx}(B)$ by subtracting a smooth background, at various temperatures (T). Arrows label selected LL indices. (b) T dependence of the relative amplitude of SdH oscillation in $\Delta R_{xx}(B)$ for the $10^{th}$ LL. Solid line is a fit to Lifshitz-Kosevich formula, from which we extract LL energy gap $\Delta E$. Inset shows $\Delta E$ as a function of B (magnetic field positions of minima in $R_{xx}$ corresponding to different LLs), and the effective mass ($m^* \sim 0.14 m_e$) is extracted from the slope of the linear fitting. (c) ln[D] (defined in the text) plotted as a function of 1/B. The Dingle temperature $T_D = 25$ K is calculated from the slope of the linear fit, corresponding to a carrier life time $\sim 5 \times 10^{-14}$ s and an effective mobility of $\sim 620$ cm$^2$/Vs.

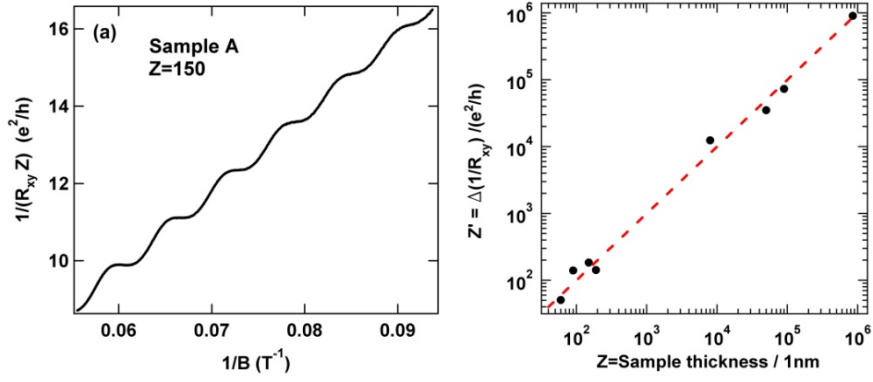

FIG. 4. (a) $1/R_{xy}$ divided by the number ($Z = 150$) of QLs plotted as a function of $1/B$, displaying plateaus separated by $\sim e^2/h$ between adjacent LLs. (b) Z', step size between the plateaus in $1/R_{xy}$ in unit of $e^2/h$, plotted against measured number (Z) of QLs for multiple samples of different thickness. The dotted line labeling Z'=Z is a guide to eyes. These samples were grown at two different sources (Purdue and Fudan Universities) with bulk carrier densities (as determined by low B Hall effect) ranging from $\sim 3\times 10^{19}$ cm$^{-3}$ to $\sim 6\times 10^{19}$ cm$^{-3}$. The observed $1/R_{xy}$ is interpreted as due to contribution from many parallel 2D conduction channels (each found to be ~2nm (2QL) in thickness, see text for more details).